\documentclass[twocolumn,aps,showpacs,preprintnumbers,byrevtex,superscriptaddress]{revtex4}
\usepackage{graphicx}
\usepackage{dcolumn}
\usepackage{bm}

\begin{document}

\title{Generation of photon-number entangled soliton pairs through
interactions}
\author{Ray-Kuang Lee}
\affiliation{Institute of Electro-Optical Engineering, National Chiao-Tung University,
Hsinchu, Taiwan}
\author{Yinchieh Lai}
\email{yclai@mail.nctu.edu.tw}
\affiliation{Institute of Electro-Optical Engineering, National Chiao-Tung University,
Hsinchu, Taiwan}
\author{Boris A. Malomed}
\email{malomed@eng.tau.ac.il}
\affiliation{Department of Interdisciplinary Studies, Faculty of Engineering, Tel Aviv
University, Tel Aviv 69978, Israel}
\date{\today}

\begin{abstract}
Two new simple schemes for generating macroscopic (many-photon)
continuous-variable entangled states by means of continuous interactions
(rather than collisions) between solitons in optical fibers are proposed.
First, quantum fluctuations around two time-separated single-component
temporal solitons are considered. Almost perfect correlation between the
photon-number fluctuations can be achieved after propagating a certain
distance, with a suitable initial separation between the solitons. The
photon-number correlation can also be achieved in a pair of vectorial
solitons with two polarization components. In the latter case, the
photon-number-entangled pulses can be easily separated by a polarization
beam splitter. These results offer novel possibilities to produce entangled
sources for quantum communication and computation.
\end{abstract}

\maketitle

\textit{Introduction} Quantum-noise squeezing and correlations are
two key quantum properties that can exhibit completely different
characteristics when compared to the predictions of the classical
theory. Almost all the proposed applications to quantum
measurements and quantum information treatment utilize either one
or both of these properties. In particular, solitons in optical
fibers have been known to serve as a platform for demonstrating
macroscopic quantum properties in optical fields, such as
quadrature squeezing, amplitude squeezing, and both intra-pulse
and inter-pulse correlations. The development of quantum theories
of nonlinear optical pulse propagation in the past years has
opened a way to analyze the quantum features of fiber-optic
solitons. Experimental progress in demonstrating various quantum
properties of these solitons has also been reported, see Refs.
\cite{Silberhorn01}-\cite{Konig02} and references therein.

In the field of quantum information processing and quantum computing,
nonlocally entangled optical quantum states have been shown to be highly
useful sources. The applications include quantum cryptography \cite{Raz99},
teleportation \cite{Bennett93}, and algorithms \cite{Shor94, Grover96}.
Following the Bohm's suggestion \cite{Bohm}, the entangled pairs were mostly
realized in terms of discrete quantum variables, such as spin, polarization,
etc. However, the original \textit{gedanken experiment} proposed by
Einstein, Podolsky, and Rosen (EPR) utilized the continuous variables (the
coordinate and momentum of a particle) to argue that the quantum mechanics
is incomplete \cite{EPR}. In 1992, Ou \textit{et al}. used nondegenerate
parametric amplification to demonstrate the EPR paradox with continuous
variables \cite{Ou92}. Later, Vaidman proposed a generalized method for the
teleportation of continuous-variable quantum states \cite{Vaidman94}.
Braunstein and Kimble analyzed the entanglement fidelity of quantum
teleportation with continuous variables \cite{Braunstein98}. Quantum
teleportation of optical coherent states was experimentally realized by
using the entanglement from squeezed states \cite{Furusawa98}. After that,
quantum-information processing with continuous variables has attracted a lot
of interest as an alternative to single-photon schemes.

In previous works, continuous-variable entangled beams have been
generated by letting two squeezed fields (squeezed vacuum states
\cite{Furusawa98}, or amplitude-squeezed fields \cite{Ralph98})
interfere through a beam splitter, which mathematically acts as
the Hadamard transformation. By utilizing the continuous EPR-like
correlations of optical beams, one can also realize quantum-key
distributions \cite{Silberhorn02} and entanglement swapping
\cite{Glockl03}. Thanks to these successful applications, squeezed
states become essential for generating entangled
continuous-variable quantum states, and play an important role in
the study of the quantum-information processing.

It has been demonstrated that two independent squeezed pulse states can be
simultaneously generated by using optical solitons in the Sagnac fiber loop
configuration \cite{Silberhorn01}. An EPR pulse source can be obtained by
combining the two output pulse squeezed states by means of a 50:50 beam
splitter. In contrast to this known method for achieving the entanglement,
in this work we propose simple schemes for generating continuous-variable
entangled states through soliton-soliton interactions, in single-mode and
bimodal (two-component) systems, \emph{without} using beam splitters. The
quantum interaction of two time-separated solitons in the same polarization
is described by the quantum Nonlinear Schr{\"{o}}dinger Equation (NLSE), and
in the bimodal system, including two polarizations, it is described by a
system of coupled NLSEs. The photon-number correlation between the two
solitons can be numerically calculated by using the \textit{back-propagation
method} \cite{Lai95}. In addition to the transient multimode correlations
induced by cross-phase modulation \cite{Konig02}, we also find nearly
maximum photon-number entanglement in the soliton pair. By controlling the
initial separation of the two solitons, one can achieve a positive quantum
correlation with the correlation parameter close to 1.

\textit{The single-mode system} Neglecting loss and higher-order effects,
which are immaterial for the experimentally relevant range of the
propagation distance $z$, temporal solitons in optical fibers are described
by the NLSE in the normalized form,
\[
iU_{z}+\frac{1}{2}U_{tt}+|U|^{2}U=0,
\]where $t$ is the retarded time \cite{Agrawal95}. The input profile of the
soliton pair is taken as
\begin{equation}
U(z,t)=\mathrm{sech}(z,t+\rho )+\gamma \,\mathrm{sech}(z,t-\rho )e^{i\theta
},  \label{initial}
\end{equation}with $\gamma $, $\theta $, and $2\rho $ being, respectively, the relative
amplitude, phase and separation of the solitons. If $\theta =0$ (the
in-phase pair), the two solitons will collide periodically in the course of
the propagation. Otherwise, they attract each other within a short
propagating distance, and then move apart due to repulsion between them.

One can evaluate the multimode quantum fluctuations around the solitons by
solving the linearized quantum NLSE \cite{Haus90}. For the case of
two-soliton collisions \cite{Konig02}, K{\"{o}}nig \textit{et al}. used an
exact classical solution for the in-phase soliton pair, and had found that
the colliding solitons carry both intra-pulse and inter-pulse photon-number
correlations. However, the photon-number correlation between the colliding
solitons is transient, i.e., the inter-pulse correlation vanishes after the
collision. Unlike the case of the collision, the photon-number correlations
caused by the \emph{continuous interaction} between the solitons belonging
to the pair initiated by the configuration (\ref{initial}) persist with the
propagation.

\begin{figure}[p]
\begin{center}
\includegraphics[width=3.0in]{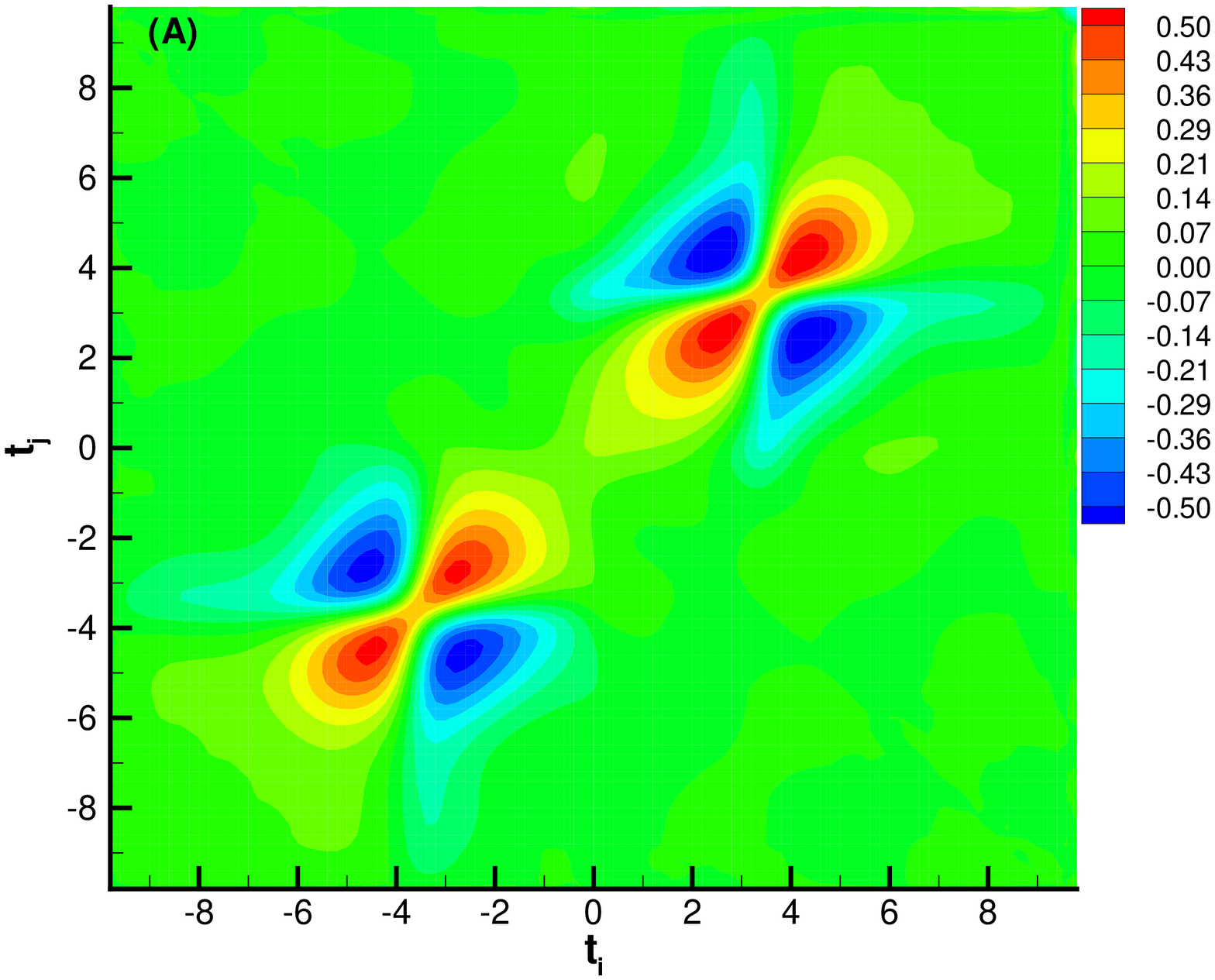}
\includegraphics[width=3.0in]{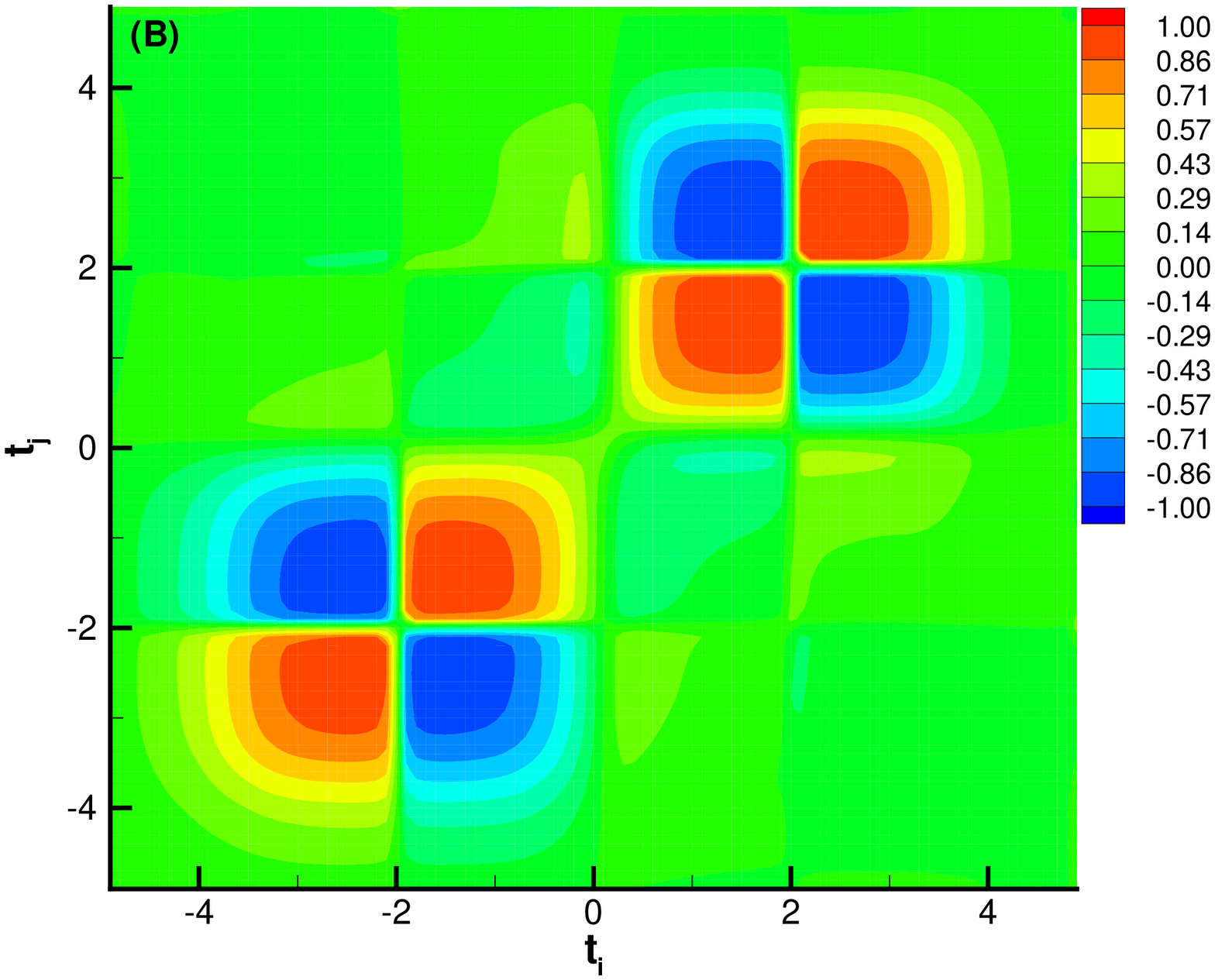}
\includegraphics[width=3.0in]{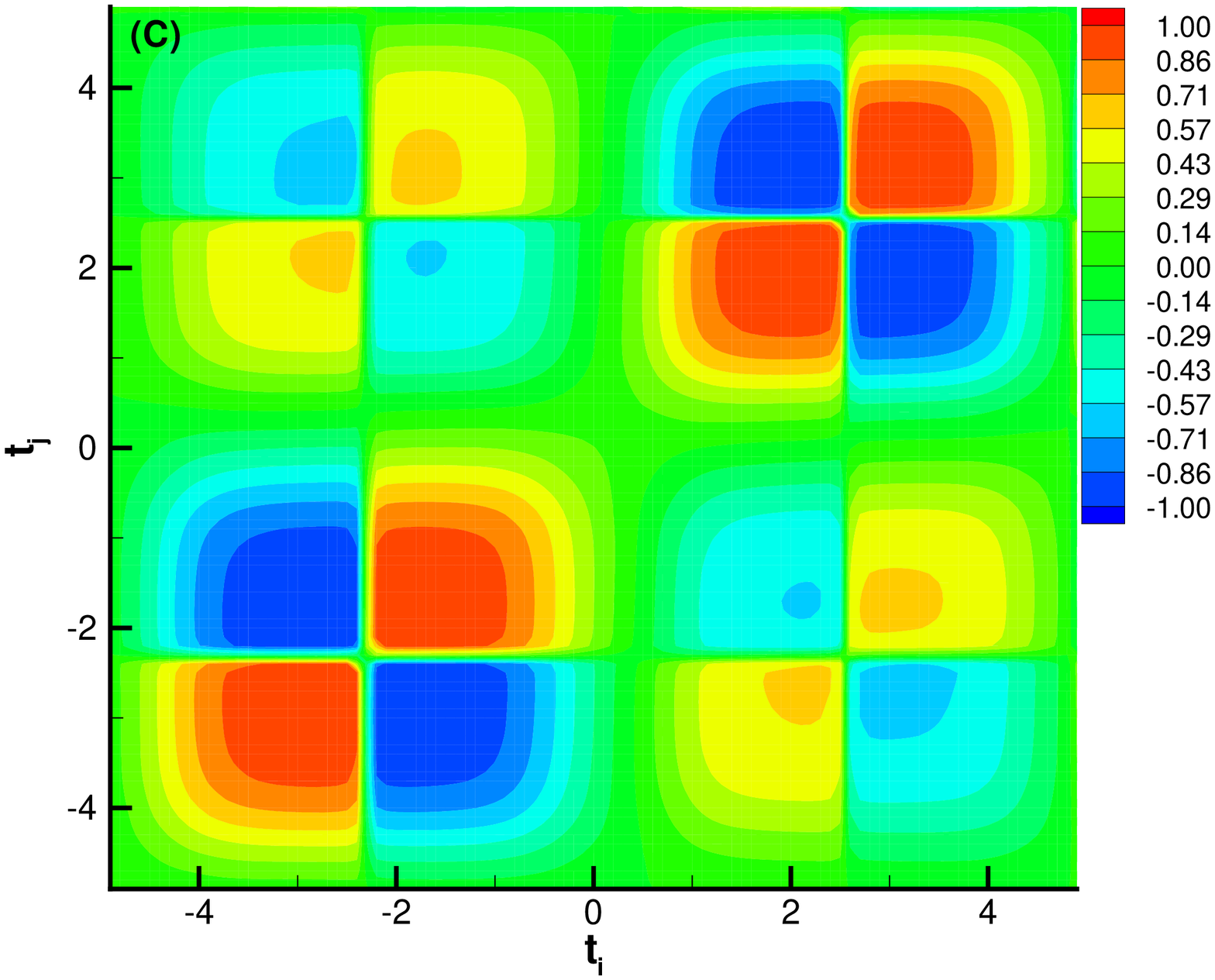}
\end{center}
\caption{The pattern of time-domain photon-number correlations,
$C_{ij}$, of two interacting out-of-phase solitons, with
$\protect\theta =\protect\pi /2$, $\protect\rho =3.5$, and
$\protect\gamma =1.0$ in Eq. (\protect\ref{initial}). The
propagation distance is $z=6$ (a), $30$ (b), and $50$ (c), in the
normalized units. The width of the time slots is $\Delta t=0.1$.
Note the difference in the bar-code scales in the panels (a) and
(b), (c).} \label{f-spectra}
\end{figure}

In Fig. \ref{f-spectra}, we display the result of evaluation of
the time-domain photon-number correlations for the out-of-phase
($\theta =\pi /2$) two-soliton pair. The correlation coefficients,
which are defined through the normally ordered covariance,

\begin{equation}
C_{ij}\equiv \frac{\langle :\Delta \hat{n}_{i}\Delta
\hat{n}_{j}:\rangle }{\sqrt{\Delta \hat{n}_{i}^{2}\Delta
\hat{n}_{j}^{2}}}~,  \label{C}
\end{equation}were calculated by means of the above-mentioned back-propagation method \cite{Lai95}. In Eq. (\ref{C}), $\Delta \hat{n}_{j}$ is the photon-number
fluctuation in the $i$-th slot $\Delta t_{i}$ in the time domain,
\[
\Delta \hat{n}_{i}=\int_{\Delta t_{i}}d\,t[U(z,t)\Delta \hat{U}^{\dag
}(z,t)+U^{\ast }(z,t)\Delta \hat{U}(z,t)],
\]where $\Delta \hat{U}(z,t)$ is the perturbation of the quantum-field
operator, $U(z,t)$ is the classical unperturbed solution, and the integral
is taken over the given time slot. As could be intuitively expected, nonzero
correlation coefficients are found solely in the diagonal region of the
spectra (\textit{intra-pulse correlations}) if the interaction distance is
short, as shown in Fig. \ref{f-spectra}(a). As the interaction distance
increases, \textit{inter-pulse correlations} between the two solitons emerge
and grow, as shown in Figs. \ref{f-spectra}(b) and (c).

In addition to the time-domain photon-number correlation pattern, we have
also calculated a photon-number \textit{correlation parameter} between the
two interaction solitons, as
\[
C_{12}=\frac{\langle :\Delta \hat{N}_{1}\Delta \hat{N}_{2}:\rangle
}{\sqrt{\Delta \hat{N}_{1}^{2}\Delta \hat{N}_{2}^{2}}}.
\]Here $\Delta \hat{N}_{1,2}$ are the perturbations of the photon-number
operator of the two soliton, the solitons being numbered (first and second)
according to their position in the time domain.

\begin{figure}[t]
\begin{center}
\includegraphics[width=3.0in]{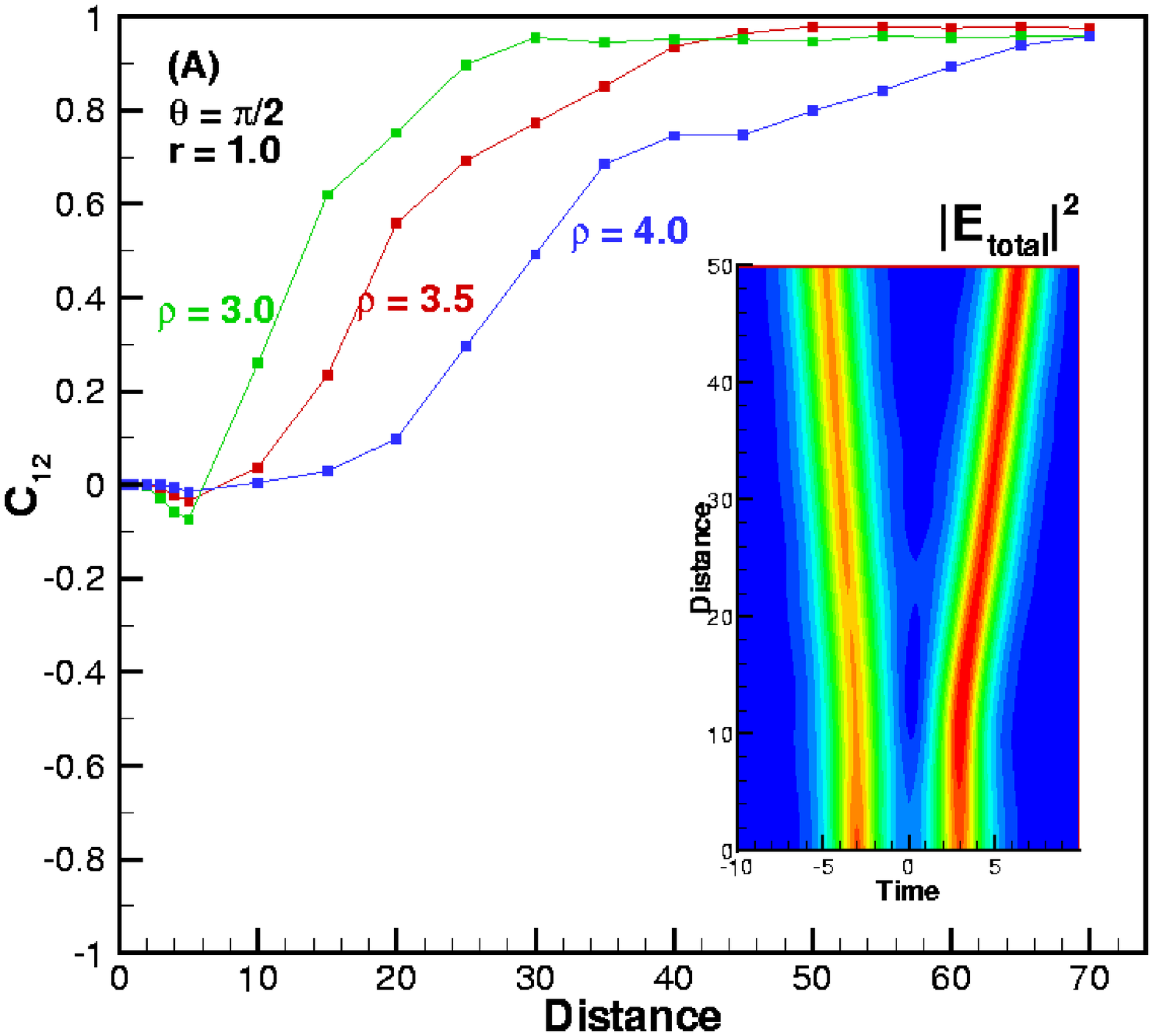}
\includegraphics[width=3.0in]{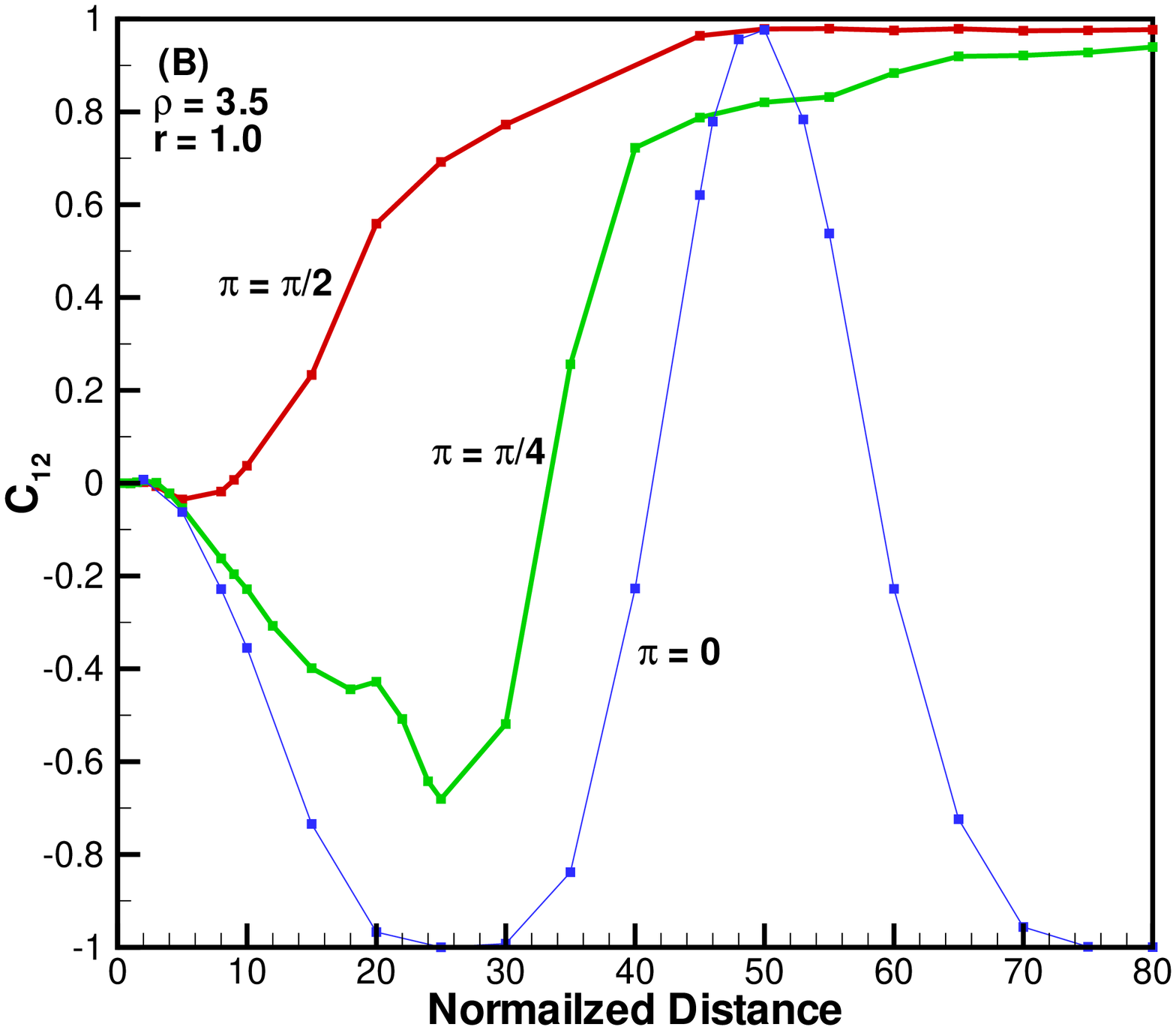}
\end{center}
\caption{The photon-number correlation parameter $C_{12}$ for the
soliton pair with different values of the separation
($\protect\rho =3.0$, $3.5$, $4.0$, while $\protect\theta
=\protect\pi /2$ and $\protect\gamma =1.0$) in (a), and different
values of the relative phase ($\protect\theta =0$, $\protect\pi
/4$, $\protect\pi /2$, while and $\protect\rho =3.5$,
$\protect\gamma =1.0$) in (b). The inset in (a) shows the
evolution of the interaction solitons by means of contour plots.}
\label{f-c12}
\end{figure}
In Fig. \ref{f-c12} (a), we show the coefficient $C_{12}$ for the soliton
pair (\ref{initial}) with the initial relative phase $\theta =\pi /2$, equal
amplitudes ($\gamma =1.0$), and different values of the separation $\rho $.
At the initial stage of the interaction, the photon-number fluctuations are
uncorrelated between the solitons, $C_{12}\approx 0$. After passing a
certain distance, the photon-number correlations between the two solitons
gradually increase, and the pair may become a nearly
maximum-positive-correlated one. The propagation distance needed to achieve
the maximum positive photon-number correlation depends on the initial
separation of the two solitons: obviously, the interaction between them is
stronger when the initial separation is smaller.

On the other hand, one can fix the initial separation but vary the initial
relative phase. For this case, the results are shown in Fig. \ref{f-c12}(b).
Similar to the case of the soliton-soliton collision \cite{Konig02}, the
photon-number correlation coefficient oscillates with the period equal to
that of the two-soliton breather, if the solitons are, initially, in-phase.
Note that in the case which may be regarded as intermediate between the
in-phase and out-of-phase ones, $\theta =\pi /4$, the correlation
coefficient first becomes negative, and then positive.

Unless $\theta =0$ (when the two solitons form a quasi-bound state in the
form of a breather), the two solitons belonging to the initial configuration
(\ref{initial}) will separate as a result of the propagation. Therefore, the
interaction between them eventually vanishes, and thus the photon-number
correlation coefficient may saturate before it has a chance to reach the
value corresponding to the total positive correlation, which is clearly seen
in the inset to Fig. \ref{f-c12}(a).

\textit{The bimodal system} The time-division entangled soliton pair
considered above can be separated with an optical switch. Since the time
separation between the two solitons is, typically, on the order of a few
picoseconds, a lossless ultrafast optical switch will be required for the
actual implementation of the scheme. The experimental difficulties can be
greatly reduced if another scheme is used, which utilizes vectorial solitons
in two polarizations. The model is based on the well-known system of coupled
NLSEs \cite{Agrawal95},
\begin{eqnarray}
i\frac{\partial U}{\partial z}+\frac{1}{2}\frac{\partial ^{2}U}{\partial
t^{2}}+A|U|^{2}U+B|V|^{2}U &=&0,  \label{eq_vectorU} \\
i\frac{\partial V}{\partial z}+\frac{1}{2}\frac{\partial ^{2}V}{\partial
t^{2}}+A|V|^{2}V+B|U|^{2}V &=&0.  \label{eq_vectorV}
\end{eqnarray}Here $U$ and $V$ are the fields in orthogonal circular polarizations, $A$
and $B$ being the self-phase- and cross-phase-modulation coefficients,
respectively, with the relation $A:B$ as $1:2$ in the ordinary optical
fibers \cite{Agrawal95}. We take the following initial configuration for the
soliton pair [cf. Eq. (\ref{initial})],
\[
U=\mathrm{sech}(t+t_{1})+\mathrm{sech}(t-t_{1}),V=\mathrm{sech}(t+t_{1})-\mathrm{sech}(t-t_{1}).
\]

\begin{figure}[tbp]
\begin{center}
\includegraphics[width=3.0in]{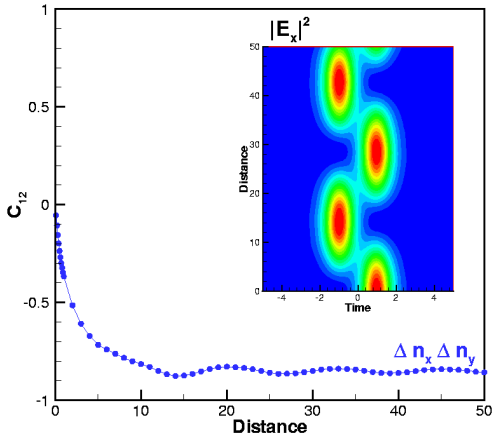}
\end{center}
\caption{The photon-number correlation coefficient of interacting vectorial
solitons. The inset displays the evolution of the $x$-component of the
classical field.}
\label{f-bound}
\end{figure}

Using the methods for the analysis of the classical vectorial
solitons developed in Refs. \cite{Haelterman93, Kaup93,
Malomed98}, we calculated the respective quantum fluctuations and
the photon-number correlators numerically. It should be noted the
total intensity of the vectorial solitons, defined in terms of the
circular polarizations, remain unchanged during the propagation,
but the intensities of the linearly-polarized ($x$- and $y$-)
components, $E_{x}=(U+V)/\sqrt{2}$ and $E_{y}=(U-V)/i\sqrt{2}$,
evolve periodically, as shown in the insert of Fig. \ref{f-bound}.
In this figure, we display the evolution of the photon-number
correlation between the $x$- and $y$- components of the vectorial
solitons, which are originally uncorrelated, and then become
negatively correlated. Recently, Lantz \textit{et al}. have showed
that vectorial solitons in the \emph{spatial domain} can also
develop an almost perfect negative correlation between quantum
fluctuations around an incoherently-coupled soliton pair
\cite{Lantz03}.

\textit{Conclusion} We have studied the quantum photon-number correlations
induced by interactions between two solitons in the time-division and
polarization-division pairs. In the former case, using the pair with
suitable initial separation and relative phase, one can generate positive or
negative photon-number-correlated soliton pairs. An ultrafast optical switch
will be needed to separate the two entangled solitons into different
channels. On the other hand, by using the vectorial solitons with two
polarization components, pairs with negative photon-number correlations
between the solitons can be generated. For this case, a simple polarization
beam splitter will be sufficient to separate the two entangled solitons into
different channels.

Such new photon-number-correlated soliton pairs feature unique entanglement
properties, which may offer new possibilities for applications to quantum
communications and computation. The applications will be considered in
detail elsewhere.

\end{document}